%% file: paper.tex
\documentclass[10pt,twocolumn]{article}
\pdfoutput=1
\usepackage{cite}
\usepackage[T1]{fontenc}
\usepackage{hyperref}
\usepackage{url}
\usepackage{microtype}
\usepackage{mdframed}
\usepackage{dcolumn}
\usepackage{graphicx}
\usepackage{xcolor}
\usepackage[font=small,labelfont=bf]{caption}
\usepackage{color,soul}
\usepackage{geometry}
\hypersetup{pdfauthor={Facebook},pdftitle={Glow: Graph Lowering Compiler Techniques for Neural Networks}}

\definecolor{darkgreen}{rgb}{0.0, 0.42, 0.24}

\usepackage{listings}
\def\CC{{C\nolinebreak[4]\hspace{-.05em}\raisebox{.4ex}{\tiny\bf ++}}}

\makeatletter
\lstdefinelanguage{llvm}{
  morecomment = [l]{;},
  morestring=[b]",
  sensitive = true,
  classoffset=0,
  keywordstyle=\color{blue},
  morekeywords={
    define, declare, program, global, constant,
    internal, external, private,
    linkonce, linkonce_odr, weak, weak_odr, appending,
    common, extern_weak,
    thread_local, dllimport, dllexport,
    hidden, protected, default,
    except, deplibs,
    volatile, fastcc, coldcc, cc, ccc,
    x86_stdcallcc, x86_fastcallcc,
    ptx_kernel, ptx_device,
    signext, zeroext, inreg, sret, nounwind, noreturn,
    nocapture, byval, nest, readnone, readonly, noalias, uwtable,
    inlinehint, noinline, alwaysinline, optsize, ssp, sspreq,
    noredzone, noimplicitfloat, naked, alignstack,
    module, asm, align, tail, to,
    addrspace, section, alias, sideeffect, c, gc,
    target, datalayout, triple,
    blockaddress
  },
  classoffset=1, keywordstyle=\color{blue},
  morekeywords={
    fadd, sub, fsub, mul, fmul,
    sdiv, udiv, fdiv, srem, urem, frem,
    and, or, xor,
    icmp, fcmp,
    eq, ne, ugt, uge, ult, ule, sgt, sge, slt, sle,
    oeq, ogt, oge, olt, ole, one, ord, ueq, ugt, uge,
    ult, ule, une, uno,
    nuw, nsw, exact, inbounds,
    phi, call, select, shl, lshr, ashr, va_arg,
    trunc, zext, sext,
    fptrunc, fpext, fptoui, fptosi, uitofp, sitofp,
    ptrtoint, inttoptr, bitcast,
    ret, br, indirectbr, switch, invoke, unwind, unreachable,
    malloc, alloca, free, load, store, getelementptr,
    extractelement, insertelement, shufflevector,
    extractvalue, insertvalue,
  },
  classoffset=2, keywordstyle=\color{purple},
  alsoletter={\%},
  keywordsprefix={\%},
}
\makeatother

\makeatletter
\lstdefinelanguage
   [x64]{Assembler}     % add a "x64" dialect of Assembler
   [x86masm]{Assembler} % based on the "x86masm" dialect
   {morekeywords={CDQE,CQO,CMPSQ,CMPXCHG16B,JRCXZ,LODSQ,MOVSXD, %
                  POPFQ,PUSHFQ,SCASQ,STOSQ,IRETQ,RDTSCP,SWAPGS, %
                  rax,rdx,rcx,rbx,rsi,rdi,rsp,rbp, %
                  r8,r8d,r8w,r8b,r9,r9d,r9w,r9b, %
                  r10,r10d,r10w,r10b,r11,r11d,r11w,r11b, %
                  r12,r12d,r12w,r12b,r13,r13d,r13w,r13b, %
                  r14,r14d,r14w,r14b,r15,r15d,r15w,r15b,ymm0,ymm1,ymm2,ymm3,ymm4}} % etc.
\makeatother

\definecolor{header-color}{gray}{0.94}
\newcolumntype{L}[1]{>{
  \raggedright\let\newline\\\arraybackslash\hspace{0pt}}m{#1}}
\newcolumntype{C}[1]{>{
    \centering\let\newline\\\arraybackslash\hspace{0pt}}m{#1}}

\begin{document}

\title{\vspace{-1em}\huge{\textbf{Glow: Graph Lowering Compiler Techniques for Neural Networks}}}

\author{%
  \parbox{\linewidth}{\centering
  Nadav Rotem,
  Jordan Fix,
  Saleem Abdulrasool,
  Garret Catron,
  Summer Deng,\\
  Roman Dzhabarov,
  Nick Gibson,
  James Hegeman,
  Meghan Lele,
  Roman Levenstein,\\
  Jack Montgomery,
  Bert Maher,
  Satish Nadathur,
  Jakob Olesen,
  Jongsoo Park,\\
  Artem Rakhov,
  Misha Smelyanskiy,
  Man Wang\\
  Facebook
  }
}%

\date{}
\maketitle

\input{body}

{\footnotesize
\bibliography{paper}}
\bibliographystyle{unsrt}

\end{document}

%% file: body.tex
\begin{abstract}
This paper presents the design of Glow, a machine learning compiler for
heterogeneous hardware. It is a pragmatic approach to compilation that enables
the generation of highly optimized code for multiple targets. Glow lowers the
traditional neural network dataflow graph into a two-phase strongly-typed
intermediate representation. The high-level intermediate representation allows
the optimizer to perform domain-specific optimizations. The lower-level
instruction-based address-only intermediate representation allows the compiler
to perform memory-related optimizations, such as instruction scheduling, static
memory allocation and copy elimination. At the lowest level, the optimizer
performs machine-specific code generation to take advantage of specialized
hardware features. Glow features a lowering phase which enables the compiler to
support a high number of input operators as well as a large number of hardware
targets by eliminating the need to implement all operators on all targets. The
lowering phase is designed to reduce the input space and allow new hardware
backends to focus on a small number of linear algebra primitives.
\end{abstract}

\section{Introduction}
\label{sec:intro}

The end of power saving due to Moore's Law, combined with the increased demand
for compute power driven by machine learning, has led to a wave of innovation in
computer architecture. Hennessy and Patterson~\cite{hp} present five principles
that guide the design of machine-learning domain specific architectures (DSA):
dedicated local memories, large numbers of arithmetic units, simple forms of
parallelism, reduced bit-widths, and domain-specific programming
models. Compilers need to perform advance whole-graph optimizations in order to
execute neural networks efficiently on DSAs. This paper describes some of these
techniques as implemented in Glow, an open-source machine learning compiler
framework for heterogeneous hardware.

Traditional machine learning frameworks iterate over the nodes in the graph and
execute them one by one. Unfortunately this node-visitor method of execution is
inefficient, even on traditional processors. As a result, machine learning
frameworks have started to hand over the graph to compilers~\cite{xla, tvm} that
execute code more efficiently. Based on the increasing importance of neural
networks, the need for energy efficiency in data centers and mobile devices, and
the design principles of domain-specific architectures, we believe that the
machine learning frameworks of the future will focus on providing attractive
programming models on top of a layer that integrates compilers for many
different targets.

In Glow, we focus on the lower parts of the software stack. We work to provide
PyTorch~\cite{pytorch} and other frameworks with a low-level graph and a code
generator for neural networks. The name Glow is an abbreviation for
Graph-Lowering, which is the main technique that the compiler uses for
generating efficient code. The Glow low-level graph will not replace the machine
learning high-level graph, in the same way that the low-level intermediate
representation in compilers does not replace the abstract syntax tree.

We aim to provide a useful compiler toolkit that will allow hardware developers
to focus on implementing efficient acceleration hardware, each of which likely
differ in capabilities, and use Glow for automating compilation tasks such as
instruction selection, memory allocation and graph scheduling. The full compiler
toolkit is open-source and publicly
available\footnote{\url{http://github.com/pytorch/glow}}.

Cadence\textsuperscript{\sffamily{\textregistered}},
Esperanto\textsuperscript{\sffamily{\textregistered}},
Habana\textsuperscript{\sffamily{\textregistered}},
Intel\textsuperscript{\sffamily{\textregistered}}, and Qualcomm Technologies
Inc.\textsuperscript{\sffamily{\textregistered}}, a subsidiary of Qualcomm
Incorporated\textsuperscript{\sffamily{\textregistered}}, have committed to
supporting Glow in future silicon products. Each of their accelerators will
likely differ in capabilities, and can use Glow for automating compilation tasks
such as instruction selection, memory allocation, and graph scheduling.

\section{Related Work}
\label{sec:related}

\subsection{Relationship to Neural Network Frameworks}
\label{sec:related_frameworks}

Frameworks such as PyTorch~\cite{pytorch}, Caffe~\cite{caffe}, and
TensorFlow~\cite{tensorflow} have found success by providing a useful way for
developers to create neural network models, and executing them on specific
architectures. However, supporting new architectures and operators is not
scalable, because adding a new operator requires it to be implemented on each
supported architecture, and adding a new architecture requires all operators be
implemented for it. Glow is designed to consume a neural network compute graph,
optimize it, and code generate for it for a diverse set of backends in a more
scalable way. This includes target-independent optimizations and analysis prior
to efficiently targeting a specific backend.

ONNX~\cite{onnx} is an open-source format for representing and serializing AI
models. It allows for interoperability between different AI frameworks, allowing
compute graphs from one framework such as PyTorch to be converted to or from
another framework such as Cognitive Toolkit (CNTK)~\cite{cntk}.

\subsection{Compiler-Related Projects}
\label{sec:related_compilers}

Several prior systems use a compiler-oriented approach to optimizing neural
networks. TensorFlow's XLA~\cite{xla} compiles neural networks for CPUs, GPUs
and accelerators. It is a practical compiler that solves actual problems. XLA is
used in production to drive a massive fleet of accelerators at Google. It lowers
nodes into primitive linear algebra operations, and then calls into a
backend-specific library for different backend (such as Eigen~\cite{Eigen} for
CPUs, or cuDNN~\cite{cudnn} for GPUs) to perform the bulk of computation. We
point out that XLA emits vectorized LLVM intermediate representation
(IR)~\cite{llvm} for some nodes (such as dot), and relies on the LLVM
vectorizer~\cite{vectorization} for other nodes. It aims to provide a backend
flexibility for TensorFlow, in a similar way that Glow is working toward
providing for PyTorch and other neural network frameworks.

TVM/NNVM~\cite{nnvm, tvm} lowers nodes into a low-level Halide-based IR wherein
loop-based optimizations can be performed. Halide~\cite{halide} is then used to
generate LLVM or CUDA/Metal/OpenCL source code. On the other hand,
DLVM~\cite{dlvm} lowers DLVM IR into LLVM IR, benefiting from the LLVM's mature
optimizer and code generator.

Yet another approach is taken by nGraph~\cite{ngraph}, which consumes a
framework's (such as Tensorflow) compute graph to represent internally in a
single level IR, and then lowers that to different backends such as cuDNN and
MKL-DNN~\cite{mkl}.

Finally, Tensor Comprehensions~\cite{tc} provides a language for neural network
developers to specify their networks such that a JIT compiler can
algorithmically search for the most efficient execution plan possible. This
execution plan is then generated in a language suited for a specific backend,
such as CUDA for a GPU, and compiled by a compiler for that language. Tensor
Comprehensions is a good solution for programmers that seek to create new
operators that do not exist today and execute them efficiently.

Similar to Glow, these systems include one or more levels of IR
(Section~\ref{sec:ir}) which represent the compute graph of some neural network
model. Additionally, like Glow, many represent tensors as first-class members
with a shape and an element type. Glow uses multiple levels of its own IR in the
entire stack, and leaves it up to each backend to implement further lowering if
desired. For example, Glow's CPU backend executes low-level Glow instructions
and calls into its own libjit standard library kernels implemented in \CC{} and
compiled with LLVM (Section~\ref{sec:cpu}).

\section{Intermediate Representation}
\label{sec:ir}

\subsection{Motivation}
\label{sec:ir_motivation}

\begin{figure}[tb]
  \begin{lstlisting}[language=C++]
for (...) A[i] = 3;
for (...) A[i] = 4;
return A[0];
  \end{lstlisting}
\vspace{-12pt}
\caption{Compilers struggle to analyze and optimize this code when the two loops
  come from two different nodes in the dataflow graph.}
\label{fig:hard_to_analyze}
\end{figure}

In this section we describe the motivation for having a high-level intermediate
representation (IR). Neural networks are dataflow graphs where each operation
represents some mathematical operation, such as element-wise add or matrix
multiplication. One way to compile this graph into an executable would be to
translate each mathematical operation directly into some low-level compiler IR
that contains loops and other low-level instructions. However, we believe that
the high-level domain specific intermediate representation is necessary for
optimizing the graph.

Consider the code in Figure~\ref{fig:hard_to_analyze}. Two for-loops write into
some memory region, and later the return-statement reads from some element in
the array. Neither GCC nor LLVM were able to remove the first loop, which is
redundant, or replace the load operation with the constant value `4'. The reason
is that analyzing loops and memory is difficult. The compiler needs to prove
that the indices in the loops do not overflow, and that pointers in the program
do not alias, and that the result of the computation is accurate and conform
with the specification of the C programming language.

Thus, trusting a sufficiently good \CC{} compiler to optimize neural networks is
not a viable strategy, because for example it is difficult to reverse engineer a
sequence of 7 loops into a convolution. Instead we have implemented a high-level
intermediate representation that allows a compiler to reason about and optimize
high-level constructs such as tensors and operations.

Glow is a retargetable compiler that supports a number of different
backends. This means that the first few phases of the compiler are
target-independent, but as you get closer to instruction selection the IR
becomes more target-specific. This design is not unique to Glow. Many compilers
and virtual machines use similar techniques to gradually canonicalize, optimize
and lower programs into instruction streams. The first two levels of IR are
shared between all compilation targets. Compiler backends may implement
additional levels of intermediate representations.

\subsection{High-Level IR}
\label{sec:ir_high_level}

The high-level IR is a dataflow node-based graph representation that is similar
to the graph that you may find inside Caffe. When we load a neural network model
we construct this graph with a direct translation of one operator to one or more
nodes. The high-level IR is a simple graph that allows basic transformations,
such as replacing all uses of some node with another node, or modifying the
content of input tensors known at compile time (e.g. pre-trained weights). The
graph is strongly typed, which means that inputs and output have a known tensor
type (consisting of the tensor's shape and element type), and that the types of
nodes are verified by the compiler. For example, the element-wise add
instruction must operate on operands of the same type.

Some strongly-typed programming languages represent dynamic types at runtime in
a safe way. Swift~\cite{Swift} generics are an example for such type system that
allows compilation for unknown yet constrained types. We have considered the
idea of developing some kind of parametric tensor types to support features such
as varying batch sizes. However, we have decided to implement a simple strict
type system instead and let the high-level machine learning framework specialize
the computation before constructing the Glow graph. We evaluated the mechanisms
that the modern programming languages use to implement generics and concluded
that most hardware accelerators do not support some of these
mechanisms. Production systems that use Glow may generate multiple Glow graphs
for different batch sizes, or recompute the graph just-in-time.

The Glow graph is structured as a module that contains multiple functions that
contain multiple nodes. Storage nodes, which are similar to global variables in
C programs, are owned by the module and accessible to all functions of the same
module. All other nodes are owned by functions and represent the different
operations of a neural network. For example, Convolution, MaxPool,
MatrixMultiply, and so on are represented as nodes. These nodes are able to
reference and access storage nodes that are owned by their containing module.

Storage nodes are the base class for, and are implemented as, Constant and
Placeholder nodes. Constant nodes are backed by a concrete, known tensor at
compilation time. Thus, the optimizer can inspect and optimize them as it sees
fit. For example, the optimizer is able to delete unused Constant nodes,
transpose them, quantize them (Section~\ref{sec:quantization}), perform constant
propagation, etc. An example of a Constant node is the pre-trained weight input
to a Convolution node during inference.

Placeholder nodes are symbolic nodes that may have backing tensors assigned or
changed after compilation. This means that unlike Constant nodes, the optimizer
cannot inspect or optimize the contents of Placeholder nodes. If the same
function is compiled using different backing tensors bound to the Placeholder
nodes, the semantics of the program are unchanged. Inputs and outputs of Glow
programs should be modeled using Placeholder nodes. An example of a Placeholder
node is the input image data tensor for image classification in a convolutional
neural network; this input image data tensor can be changed without recompiling
the function.

As an end-to-end example, one module could contain both an inference function
and the gradient of that inference function. The weights (to be trained by the
gradient function) would be created as Placeholder nodes, as they are not
constant during training. Executing the gradient function would update the
tensors backing the weight Placeholder nodes. The nodes in the inference
function could then reference and access the weight Placeholder nodes (backed by
the now-trained weights tensors). Thus, the Placeholder nodes can be converted
to Constant nodes, and the inference function can be better optimized during
compilation, knowing these nodes are Constant.

\begin{figure}[tb]
\includegraphics[width=\columnwidth]{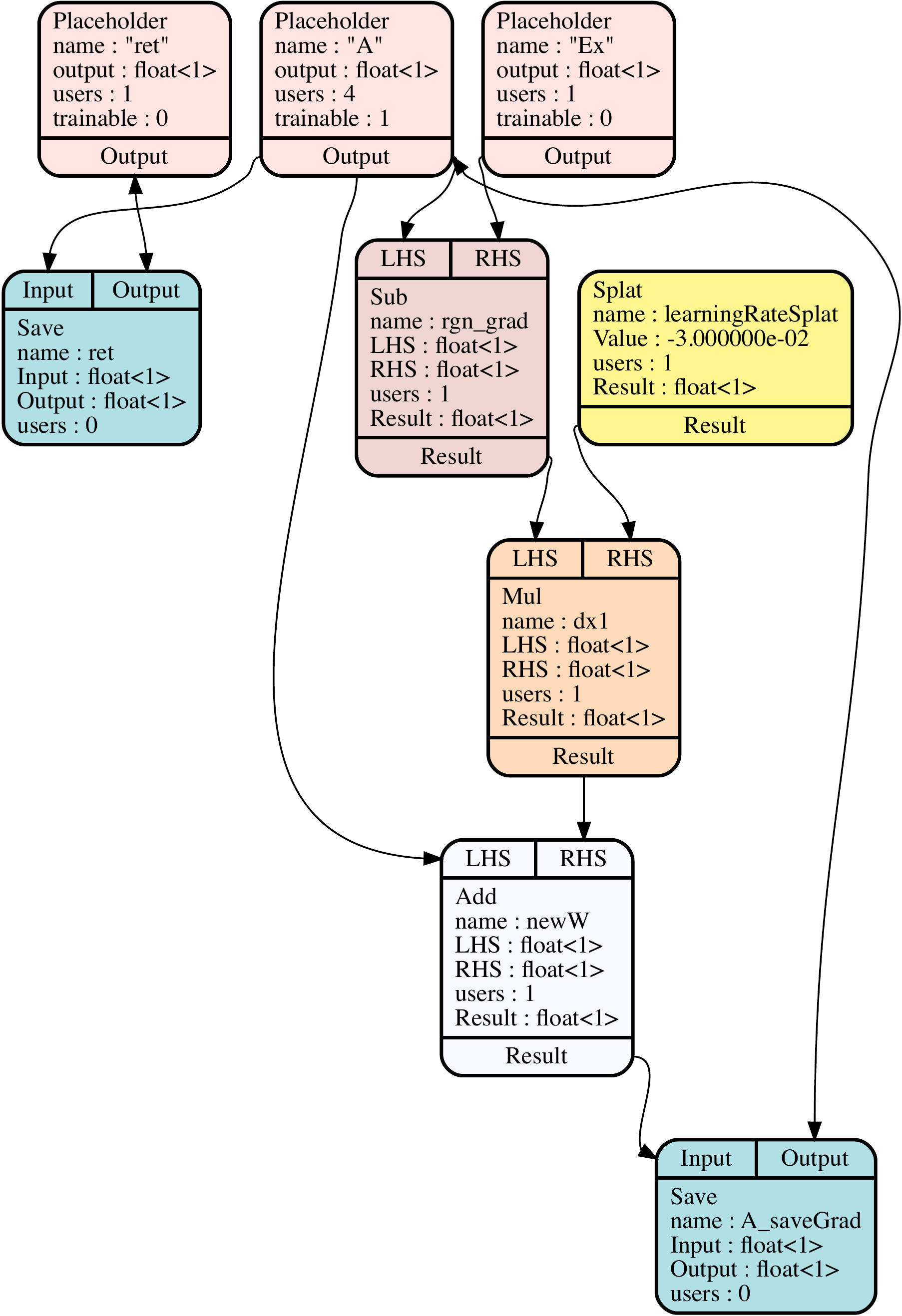}
\caption{A lowered compute graph in Glow's high-level IR, representing a
  regression of $A$, automatically differentiated by Glow.}
\label{fig:base_graph}
\end{figure}

The compiler has a debug method for dumping textual and graphical
representations of the graph. Figure~\ref{fig:base_graph} depicts the compute
graph of a regression of $A$ automatically differentiated by Glow, with the
value of Placeholder node A updated with the gradient of the expression. Glow
lowers the nodes that compute the gradient of the expression and the stochastic
gradient descent (SGD) node into a sequence of low-level operators (Sub, Mul,
Add, and Save). The different compiler backends do not need to implement support
for the DivGrad or SGD nodes.

By contrast, classic machine learning frameworks that are not able to
automatically generate fused kernels (Section~\ref{sec:stacking}) need to
implement hundreds of CUDA and CPU compute kernels that represent the un-lowered
operators. This limits their ability to support new kinds of hardware and ties
them to one or two major hardware vendors.

\subsection{Predication}
\label{sec:ir_predication}

Predication is a well-known technique to control the execution of some node or
instruction by means of a boolean flag. If the value of the flag at runtime is
set to `false' then the predicated node or instructions may return any value. A
correct program should know to ignore the output of the predicated instruction
because it could be zeros or uninitialized memory. The type of the flag must be
a boolean value or a vector of booleans that matches the batch size. Predicates
could accelerate the performance of some networks by avoiding some
computation. It can particularly be useful when applied to Recurrent Neural
Networks~\cite{lstm}, because different elements of the batch may have different
lengths and do not need to perform the same amount of computation.

\subsection{Node Lowering}
\label{sec:lowering}

The Glow compilation pipeline solves the problem of targeting a large number of
opcodes to many different targets. Modern machine learning frameworks support
hundreds of operators on many different hardware backends. The approach that is
taken by classic machine learning frameworks is to implement each opcode for
each hardware target. In such frameworks, ReLU would be implemented once for the
GPU, once for the CPU, once for some mobile DSP accelerator, and so on. This
approach does not scale as the number of opcodes and the number of hardware
targets increase.

Instead, Glow takes a different approach. Instead of compiling the high-level
operators directly, Glow performs "node lowering". In this phase, the compiler
breaks the high-level operator nodes into low-level linear algebra operator
nodes. For example, the FullyConnected layer is represented as a matrix
multiplication followed by broadcasted add. Different compiler backends do not
have to implement the FullyConnected layer and a dozen other high-level opcodes,
just the low-level matrix multiplication.

This lowering phase drives many of the design decisions of the compiler. In
Glow, lowering is performed as part of the high-level graph as described above,
prior to moving to low-level IR (Section~\ref{sec:ir_low}). This is due to a
number of reasons. First, the new lowered graph may allow for additional
graph-level optimizations. Second, the new graph structure may affect the
decisions of the instruction scheduler. And third, after lowering we allow the
backends to perform additional target-specific optimizations on the lowered
graph.

The lowering phase comes after the graph is differentiated. Because the lowering
transformation does not preserve the semantics of the graph, it is not possible
to differentiate the graph for certain operators. For example, the Regression
node (which produces gradient when optimizing total squared error) becomes a
no-op for the inference case, but is translated into an element-wise subtract
for the training case. Performing the lowering before differentiation would
prevent us from performing the correct lowering of the Regression node.

\subsection{Low-Level IR}
\label{sec:ir_low}

After optimizing the graph with target-independent optimizations, and lowering
from high-level operator nodes to linear algebra operator nodes, the code is
further lowered into the low-level IR in a phase that is called "IRGen" (which
stands for IR generation)\footnote{IRGen is optional; backends can it if they
  have their own software stack that prefers to consume the Node representation
  of the program.}. This is a one-to-many translation where each high-level node
is translated into one or more instructions.

The low-level IR enables a different kind of target independent optimizations
that are not possible with the high-level graph format. This is an
instruction-based representation that operates on tensors that are referenced by
address. This gives the compiler the ability to perform low-level memory
optimizations that are not possible at the high-level, because memory is not
represented directly. An example of such a transformation is the optimization
that allows certain operations to transform some buffers in-place, such as
element-wise arithmetic.

In the context of hardware acceleration, the low-level instruction-based
representation allows the compiler to represent device-specific operations such
as asynchronous DMA operations. Hiding the latency of memory operations is
important for utilizing the execution units of the hardware effectively, and the
instruction-based representation allows the compiler to create a schedule that
hides the latency of the memory operations.

The IR is strongly typed and each instruction operand kind has known parameter
types. It is designed to be used as an in-memory form, though can be dumped to
human readable assembly-like format.

A function in IR form contains two sections: `declare' and `program'. In the
first section of the IR we declare a number of memory regions that live
throughout the lifetime of the program. This is similar to global variables in
C. The second part of the IR is a list of instructions. Each variable is
annotated with the kind of initialization that the program should do.

There are two kinds of memory regions which correspond to these two sections:
global memory regions (found in `declare') and locally allocated regions (found
in `program'). The locally allocated memory regions are similar to `alloca' in
LLVM
IR\footnote{\url{http://llvm.org/docs/LangRef.html\#alloca-instruction}}. Memory
regions are strongly typed, which means that the kind of type of tensor that the
region represents is known.

Instructions operate on either these global memory regions or locally allocated
regions. Each operand is annotated with one of the qualifiers
`@in'/`@out'/`@inout'. `@in' means that the buffer is read from. `@out' means
that the buffer is written into. And `@inout' means that the instruction may
read and write into the buffer. These operand qualifiers help the optimizer
decide when it is legal to perform certain optimizations, such as copy
elimination or buffer sharing. Instructions may have other attributes that
specify the legality of some optimizations. For example, some instructions
require that the data from the forward pass would be kept around for the
backward pass, so if the program is not optimized for inference-only mode then
certain memory optimizations cannot happen.

\begin{figure}[tb]
  \begin{minipage}[]{\columnwidth}
    \footnotesize{
      \begin{lstlisting}[language=llvm,emph={convolution,weight,alloc,max0,pool,max,dealloc},emphstyle={\color{red}}]
declare {
 %input = weight float<8 x 28 x 28 x 1>, broadcast, 0.0
 %filter = weight float<16 x 5 x 5 x 1>, xavier, 25.0
 %filter0 = weight float<16>, broadcast, 0.100
 %weights = weight float<10 x 144>, xavier, 144.0
 %bias = weight float<10>, broadcast, 0.100
 %selected = weight index<8 x 1>
 ...
 %result = weight float<8 x 10>
}

program {
 %allo = alloc float<8 x 28 x 28 x 16>
 %conv = convolution [5 1 2 16] @out %allo, @in %input, @in %filter3, @in %bias0
 %allo0 = alloc float<8 x 28 x 28 x 16>
 %relu = max0 @out %allo0, @in %allo
 %allo1 = alloc index<8 x 9 x 9 x 16 x 2>
 %allo2 = alloc float<8 x 9 x 9 x 16>
 %pool = pool max [3 3 0] @out %allo2, @in %allo0, @inout %allo1
 ...
 %deal6 = dealloc @out %allo6
 %deal7 = dealloc @out %allo7
 %deal8 = dealloc @out %allo8
 %deal9 = dealloc @out %allo9
}
      \end{lstlisting}
    }
  \end{minipage}
\vspace{-12pt}
\caption{Unoptimized low-level Glow IR.}
\label{fig:ir}
\end{figure}

Figure~\ref{fig:ir} shows an example of unoptimized Glow IR. Note that the
`alloc' instruction does not allocate memory; it just marks the lifetime of the
activation. The low-level memory allocator is responsible for allocating all of
the buffers into a single coalesced region.

\subsection{Summary: The Lifetime of a Glow Instruction}

This section summarizes how instructions travel from the beginning of the
compilation pipeline, and through the different levels of IR and to the
backends. This is a high-level overview of the compilation process:

\begin{enumerate}
\itemsep0em
\item{The graph is either loaded via the graph loader (from ONNX or Caffe2
  format), or constructed via the \CC{} interface.}
\item{The graph is differentiated if needed.}
\item{The graph is optimized.}
\item{Linear algebra node lowering takes place.}
\item{Additional rounds of optimizations occur, both target independent and
  target specific.}
\item{The graph is scheduled into a linear sequence of nodes that minimizes
  memory usage.}
\item{IRGen converts the low-level graph into instructions.}
\item{Low-level IR optimizations are performed.}
\item{Backend-specific optimizations and code generation are performed.}
\end{enumerate}

\subsection{ClassGen}

Glow uses automatic code generation techniques (class-gen) for defining
instructions and nodes. The purpose of the automatic code generation tools in
Glow is similar to the motivation behind LLVM's TableGen, which is to help a
human develop and maintain records of domain-specific information.

The current system is capable of generating two kinds of classes: Nodes for the
high-level IR and Instructions for the low-level IR. Figure~\ref{fig:classgen}
shows an example of the code for generating the AvgPool instruction. ClassGen
generates most of the methods that instructions need to have, such as
instruction equality and hashing, cloning, printing, verification, etc.

\begin{figure}[tb]
    \begin{lstlisting}[language=C++,emph={MemberType,OperandKind,VerifyKind},emphstyle={\color{darkgreen}},basicstyle=\scriptsize\ttfamily]
BB.newInstr("AvgPool")
  .addOperand("Dest", OperandKind::Out)
  .addOperand("Src", OperandKind::In)
  .addMember(MemberType::VectorUnsigned, "Kernels")
  .addMember(MemberType::VectorUnsigned, "Strides")
  .addMember(MemberType::VectorUnsigned, "Pads")
  .autoIRGen()
  .autoVerify(VerifyKind::SameElementType,
               {"Dest", "Src"})
  .addGradientInstr({"Dest"}, {"Dest", "Src"});
    \end{lstlisting}
\vspace{-12pt}
\caption{Example class-gen for the Average Pool instruction.}
\label{fig:classgen}
\end{figure}

\section{Quantization}
\label{sec:quantization}
In the context of machine learning, quantization is the process of converting a
neural network from floating-point to integer arithmetic. Arithmetic using small
integers is more efficient than the computation of full-width floating-point
numbers, and additionally decreases memory usage.

Glow is able to convert floating-point-based networks into signed 8-bit integer
networks. The canonical quantization representation is using signed integers,
though it is possible to support other quantization formats. Glow uses
profile-guided quantization, observing execution during inference to estimate
the possible numeric range for each stage of the neural network. Training-based
quantization is considered future work.

\subsection{Tensor Representation}

In Glow, tensors are typed and can represent floats, quantized
non-floating-point values such as currently supported Int8 (8-bit signed
integers), and index types. A quantized tensor's type is made up of the
underlying element type (Int8), as well as the possible range of the values in
the tensor using `scale' and `offset' fields. To convert from the 8-bit integer
range of {[}-128..127{]} to the floating-point number that they represent, Glow
uses the conversion formula:

\begin{lstlisting}[language=C++]
     value = (input - offset) * scale
\end{lstlisting}

Activations, weights, and storage nodes all use the same type-system and
represent information in a uniform way.

\subsection{Profile-Guided Quantization}
\label{sec:quant_prof}

Different parts of the network contain floating-point values in different
ranges. In some parts, the typical range of the numbers is between zero and one,
while in other parts of the network the possible range is in the
hundreds. Choosing a single conversion scale for the whole network would not
work, because a single scale value could be imprecise for small values and
truncate large values.

\begin{figure}[tb]
\includegraphics[width=\columnwidth]{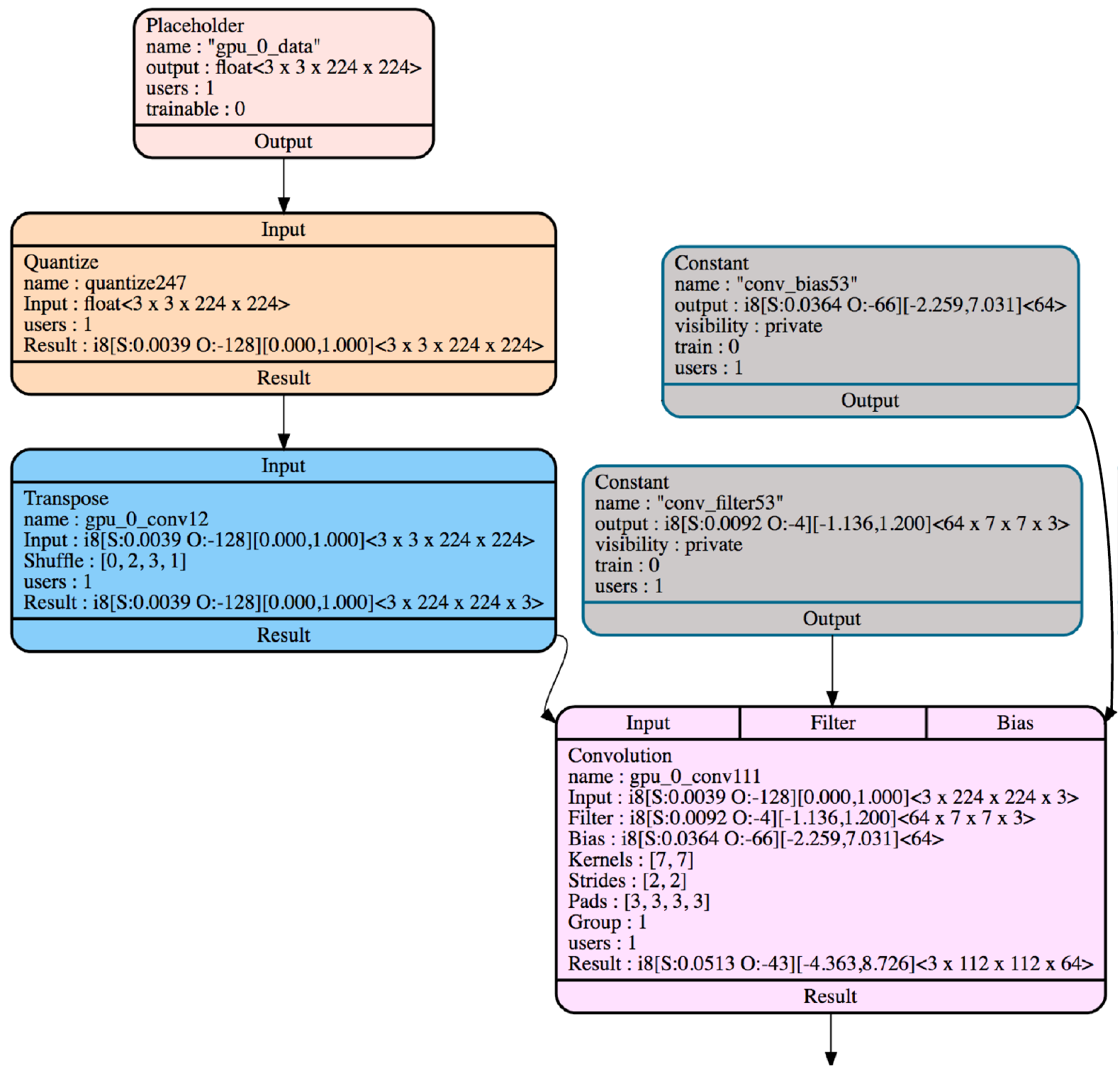}
\caption{A quantized subgraph from Resnet50.}
\label{fig:resnet50_quant}
\end{figure}

We use profile-guided information to estimate the possible numeric range for
each stage of the neural network. Our quantization conversion works using a
two-phase process. First, we statically instrument the network with special
profiling nodes that record the ranges of activations that flow in the network,
optimize the network including these profiling nodes, and then run
inference. Then, we recompile the network using this profile information to
convert the network into a quantized form, allowing for static optimization of
the quantized graph. We convert portions of the network into islands of integer
computation and aim to generate outputs in the range that the original
floating-point network produces. Figure~\ref{fig:resnet50_quant} shows a
quantized subgraph from Resnet50.

\subsection{Compiler Optimizations For Quantization}
\label{sec:quant_opt}

Glow features a number of compiler optimizations that transform the compute
graph and make it more efficient. There are a few classes of optimizations and
parameters to optimize.

First, we attempt to minimize the number of conversions between floating-point
tensors and integer tensors, in both directions. Some operations, such as
`transpose' and `concat' operate on both types, and changing the representation
can minimize conversions.

Second, the neural network contains `rescale' nodes that change the range of the
integers. These nodes are required to convert between numeric ranges that mimic
the original floating-point network. However, in many cases, it is possible to
fold the rescale operations into numeric-producing operations, and eliminate
them.

Third, it's possible to rescale the values in the network in order to allow fast
hardware implementations of the quantized operations. For example, consider the
`max' operations. By converting both sides of the `max' into the same scale we
allow the hardware to perform a simple comparison. By normalizing both sides of
the `max' operation to the same scale we enable this efficient optimization.

\section{CPU Backend}
\label{sec:cpu}

This section describes the implementation of the CPU backend. The Glow CPU
backend compiles the low-level intermediate representation into an optimized
stream of instructions. It uses LLVM to optimize and emit machine code and was
tested on x86 and ARM64. The backend can emit a stand-alone object file to disk
or execute code in just-in-time mode. The backend emits debug information, which
makes it possible to debug Glow in a debugger and place a breakpoint in specific
operator, or understand the performance of networks using a profiler.

\subsection{Standard Library}

One interesting aspect of the Glow CPU backend is the use of a small target
independent standard library. The CPU backend needs to generate code for machine
learning operators such as Convolution and SoftMax. One possibility is to call
into some external library such as Eigen. This is easy to do, and many machine
learning frameworks use this technique. The disadvantage of this technique is
that the external binary library has no information about the specific operation
that is being compiled. Some of the parameters that an optimized implementation
may care about are the specific tensor sizes, the exact addresses of buffers in
memory, and whether some pointer aliases another pointer.

Glow compiles a small standard library that ships with the compiler into LLVM
bitcode. During compilation, Glow loads the bitcode from disk and specializes
the operator implementations for the specific context. Glow replaces function
arguments that represent the dimensions of some tensor or buffer addresses with
constants that LLVM can optimize to generate efficient code. The compiler can
decide on the kind and level of operator specialization to perform, trading
compile time and binary size for performance.

Most operators are very simple and the LLVM vectorizer~\cite{vectorization} is
able to generate very efficient code. Notice that by providing the exact tensor
dimensions and loop trip count the vectorizer is able to generate efficient code
that does not contain pre-header legality check and scalar loop to handle the
remainder odd iterations. The convolution and matrix multiplication operations
are hand-optimized in \CC{} using the clang extended OpenCL vector syntax, and
LLVM does a good job allocating registers and encoding the instructions,
removing the need to use inline assembly.

\subsection{Operator Stacking}
\label{sec:stacking}

One important optimization that the CPU backend implements is stacking of
data-parallel operators. Consider a sequence of operators that operate one
element at a time, for example a ReLU, Add, Sub. Iterating over a large buffer
multiple times is inefficient because it requires the CPU to load the memory
multiple times, each time invalidating the whole cache. Instead, Glow stacks
operators and performs a few data-parallel operators one after the other on the
same memory location. Notice that as described above, this is not an
optimization that LLVM can perform by itself and it requires a special
high-level data structure.

Operator stacking is similar to operator fusion. However, when fusing multiple
operators (e.g. Conv and ReLU fused together), all backends that want to support
this fused operator must implement a specific kernel for each permutation of
operators. In contrast, Glow's stacking automatically creates such kernels; all
of the possible permutations of data-parallel nodes are automatically fused into
a fast kernel.

\label{sec:resnet50_cpu}
\begin{figure}[tb]
  \begin{lstlisting}[emph={Filter,layout,before,transformation,after},emphstyle={\color{blue}},otherkeywords={:}]
Filter layout before transformation:
   [depth, filter_x, filter_y, channel]
Filter layout after transformation:
   [depth/N, filter_x, filter_y, channel, N]
  \end{lstlisting}
\vspace{-12pt}
\caption{Transformation of a convolution's filter's memory layout to optimize
  for SIMD memory accesses. Depth refers to the output depth of the filter, and
  channel refers to the input channel.}
\label{fig:conv_transform}
\vspace{-2mm}
\end{figure}

The approach of stacking multiple operations has many advantages. First, there
is an immediate performance gain for places in the graph where data-parallel
operators are placed one on top of the other. Second, backends do not need to
implement kernels for all possible permutations of consecutive data-parallel
nodes. And lastly, it allows Glow to lower high-level operators knowing that the
backend can fuse them and recover the performance.

For example, Glow lowers the SGD (stochastic gradient descent) operator into a
sequence of low-level primitives that include addition, subtraction, and
multiplication. Lowering the SGD node into low-level primitives simplifies the
design of the compiler by reducing the operator-space that the backend needs to
handle. Operator stacking can also accelerate computation on GPUs by reducing
the kernel launch overhead.

\subsection{Use Case: Optimizing Resnet50 for the CPU}

\begin{figure}[tb]
\includegraphics[width=\columnwidth]{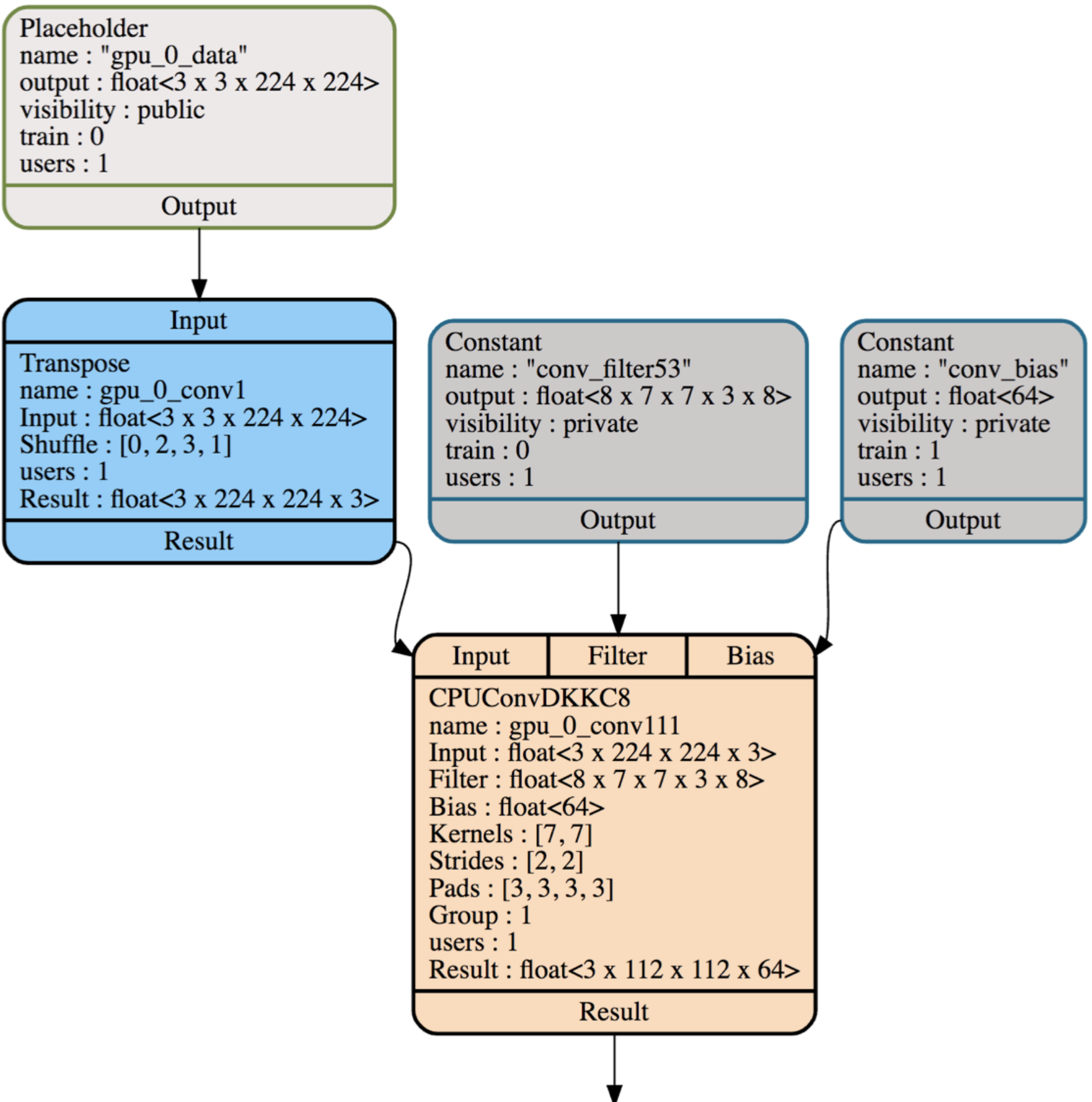}
\caption{A subgraph from Resnet50 optimized for the CPU backend. The
  CPUConvDKKC8 node requires weights with a modified memory layout for efficient
  SIMD access (Figure~\ref{fig:conv_transform}).}
\label{fig:resnet50_cpu}
\end{figure}

Here we describe the way that Glow optimizes Resnet50 to generate an efficient
stream of x86 instructions. Resnet50 is a residual convolutional neural network
containing 54 convolutions as well as other operators such as element-wise
addition, ReLU, batch normalization, max and average pooling, fully-connected,
and softmax. Glow optimizes Resnet50 by performing high-level and low-level
optimizations.

First, high-level transformations eliminate redundant transpose operations and
merge the batch normalization operation with a convolution node. Next, the CPU
backend transforms the graph into a target-specific graph that allows
device-specific optimization. The CPU backend identifies three kinds of
convolutions: convolutions with a small number of channels, convolutions where
the size of the input activation buffer is large, and convolutions where the
filter weight buffer is large. Each one of these convolutions requires a
different compilation strategy.  Next, the target-specific optimizer mutates the
graph and generates code that matches the selected convolution. Each convolution
kind uses a different filter memory layout and tile
size. Figure~\ref{fig:conv_transform} depicts the transformed filter memory
layout.

This 5-dimensional tensor layout allows for consecutive SIMD memory access. The
N parameter is selected based on the iteration order and the blocking strategy
for the convolution. The CPU backend traverses the graph and replaces any
convolutions it would like to optimize in this way with this specialized
convolution. This can be seen in Figure~\ref{fig:resnet50_cpu}.

The second parameter that the compiler controls is the size of the convolution
tile. Glow selects a processing tile that depends on the size of the first level
cache of the processor.

Next, the low-level optimizer optimizes the instruction stream by shrinking the
lifetime of memory allocations for the activations, and then performs static
memory allocation for the whole network into a single buffer. This reduces the
mutable memory footprint of the network. From this point in the compilation
pipeline the compiled code can refer to pointers in memory.

\begin{figure}[tb]
  \footnotesize{
    \begin{lstlisting}[language={[x64]Assembler},emph={vmovaps,vaddps,vmaxps,addq},emphstyle={\color{red}},otherkeywords={\%}]
LBB14_1:
  vmovaps 3211264(%rcx,%rax,4), %ymm1
  vmovaps 3211296(%rcx,%rax,4), %ymm2
  vmovaps 3211328(%rcx,%rax,4), %ymm3
  vaddps 6422528(%rcx,%rax,4), %ymm1, %ymm1
  vaddps 6422560(%rcx,%rax,4), %ymm2, %ymm2
  vmovaps 3211360(%rcx,%rax,4), %ymm4
  vaddps 6422592(%rcx,%rax,4), %ymm3, %ymm3
  vaddps 6422624(%rcx,%rax,4), %ymm4, %ymm4
  vmaxps %ymm0, %ymm1, %ymm1
  vmaxps %ymm0, %ymm2, %ymm2
  vmaxps %ymm0, %ymm3, %ymm3
  vmovaps %ymm1, 6422528(%rcx,%rax,4)
  vmovaps %ymm2, 6422560(%rcx,%rax,4)
  vmaxps %ymm0, %ymm4, %ymm1
  vmovaps %ymm3, 6422592(%rcx,%rax,4)
  vmovaps %ymm1, 6422624(%rcx,%rax,4)
  addq $32, %rax
  \end{lstlisting}
  }
\vspace{-12pt}
\caption{A loop in x86 assembly as generated by the Glow CPU Backend, with a
  fused element-wise addition and ReLU (max) operation.}
\label{fig:conv_asm}
\vspace{-4mm}
\end{figure}

Finally, the compiler performs efficient code generation for the non-convolution
parts of the network. For example, Figure~\ref{fig:conv_asm} depicts the
generated assembly for some part of the network. The compiler fused two
unrelated element-wise operations into a single loop. The Add and Max operations
are performed on the same memory buffer without reading the memory twice.

\section{Glow Runtime}
\label{sec:runtime}

\begin{figure*}[tb!]
\includegraphics[width=\linewidth]{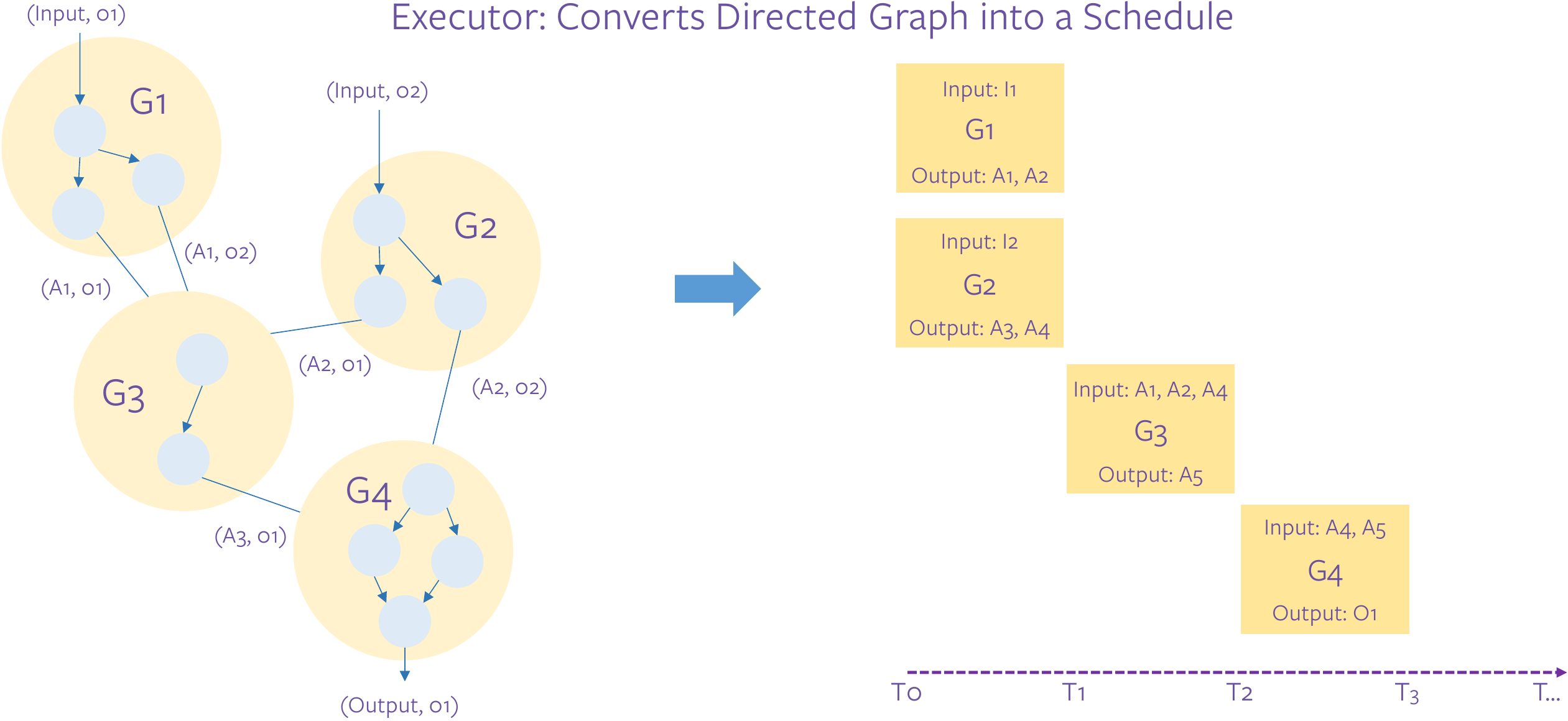}
\caption{A simple example showing a graph partitioned into multiple sub-graphs,
  themselves making up a directed graph, and then converted into a schedule by
  the executor.}
\label{fig:executor}
%% \vspace{-4mm}
\end{figure*}

\begin{figure}[tb!]
\includegraphics[width=\columnwidth]{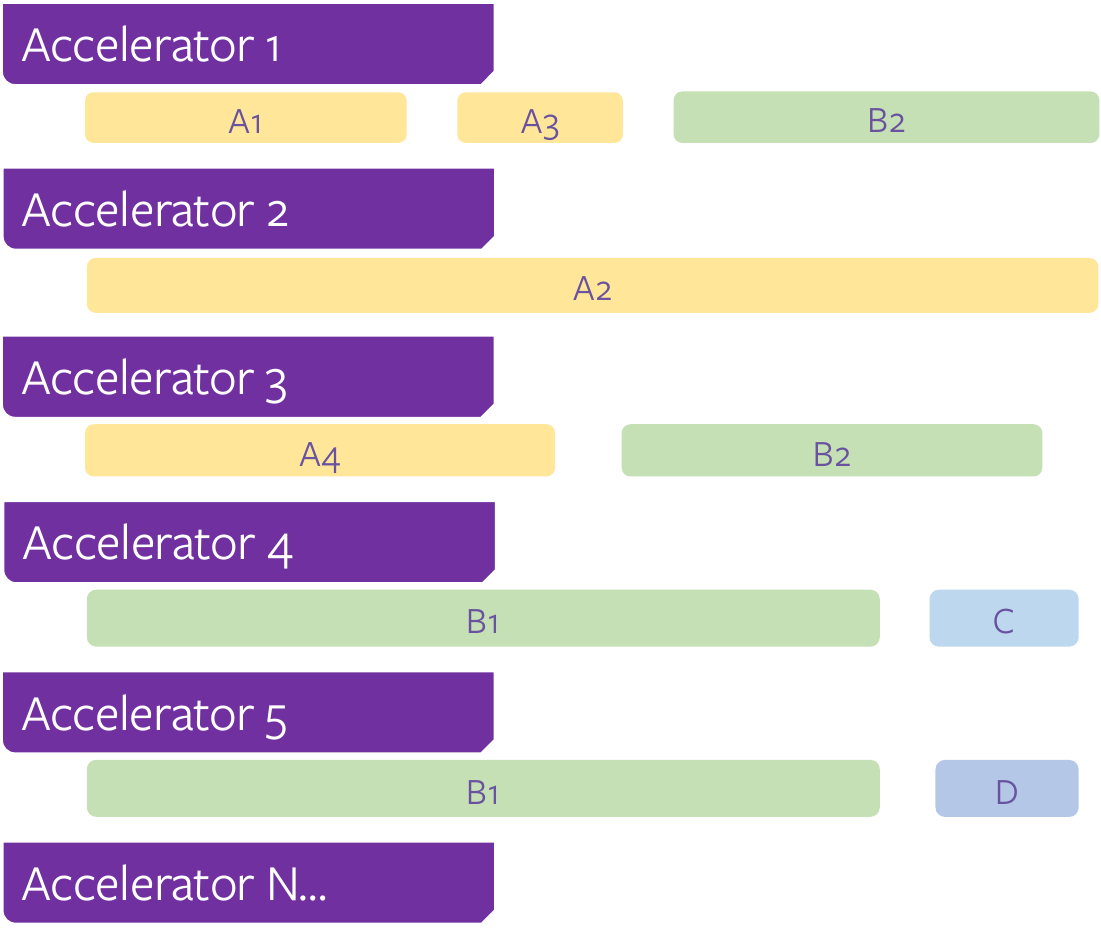}
\caption{Four networks (A, B, C, D) have been assigned to five accelerators. A
  and B have been partitioned, and B has been duplicated on multiple accelerators.}
\label{fig:provisioner}
%% \vspace{-2mm}
\end{figure}

After compiling a model, Glow provides a runtime that is capable of partitioning
models, queueing requests, and executing models across multiple devices. It
provides a host level abstraction for compiling and loading models and handling
concurrent inference requests on all those models. The runtime allows users of
Glow to target one interface without needing to worry about the underlying
hardware changing, and since it manages the host's accelerator cards, it can
take advantage of its intimate knowledge of the hardware to do the best
partitioning job possible.

\subsection{Glow Runtime Components}
The \textbf{Partitioner} splits a network into sub-networks that can be run on
multiple devices. Depending on each accelerator's available memory and the size
of the weights of a model, we may want or need to partition an input network
into sub-graphs across multiple accelerators in order to saturate each
accelerator. A network is divided into sub-networks based on different criteria:
memory constraints, estimated time cost, and communication cost between devices.

The \textbf{Provisioner} assigns partitioned sub-graphs to specific devices and
calls into the backend and Device Manager to compile and load each sub-graph
onto a device. An example can be seen in Figure~\ref{fig:provisioner}.

The \textbf{Device Manager} serves as an abstraction for the physical device. It
handles network loading, memory transfers, execution on the device, and tracks
hardware state. Just like we have a different backend per device type, there is
a DeviceManager class per device type and an instance of DeviceManager per
physical accelerator on the host.

The \textbf{Executor} handles the execution of a network. It tracks each
sub-network's execution state and propagates sub-network inputs and outputs. The
Executor is responsible for asynchronously handing incoming inference requests
for a network and returning the collated results. Figure~\ref{fig:executor}
shows a simple example of a partitioned, directed graph converted into a schedule.

\subsection{Execution Flow}

\textbf{Adding a network:}
\begin{enumerate}
\itemsep0em
\item{The Partitioner splits the network into one or more sub-networks.}
\item{The Provisioner compiles each sub-network and assigns them to one or more Devices.}
\item{One or more DeviceManagers load the sub-networks and their weights onto its associated Device.}
\end{enumerate}

\noindent
\textbf{Handling an inference request:}
\begin{enumerate}
\itemsep0em
\item{The HostManager creates a new Execution graph with intermediate storage.}
\item{The Executor kicks off the first sub-network execution.}
\item{The DeviceManager loads inputs onto the card and begins execution. When done, it reads outputs and signals completion.}
\item{The Executor triggers any sub-networks with satisfied dependencies.}
\item{When complete, the HostManager returns outputs.}
\end{enumerate}

\section{Evaluation}

We compare the performance of Glow vs. TensorFlow-1.7 and TVM. The experiments
were run on a Kaby Lake Intel\textsuperscript{\sffamily{\textregistered}} Core
i7-7600U (which does not support AVX-512) running on a single CPU core at 2.80
GHz. All three frameworks were compiled to support the native architecture. We
used the Keras library~\cite{Keras} to supply and run pre-trained models for
TensorFlow and TVM.

\begin{figure}[tb]
\includegraphics[width=\columnwidth]{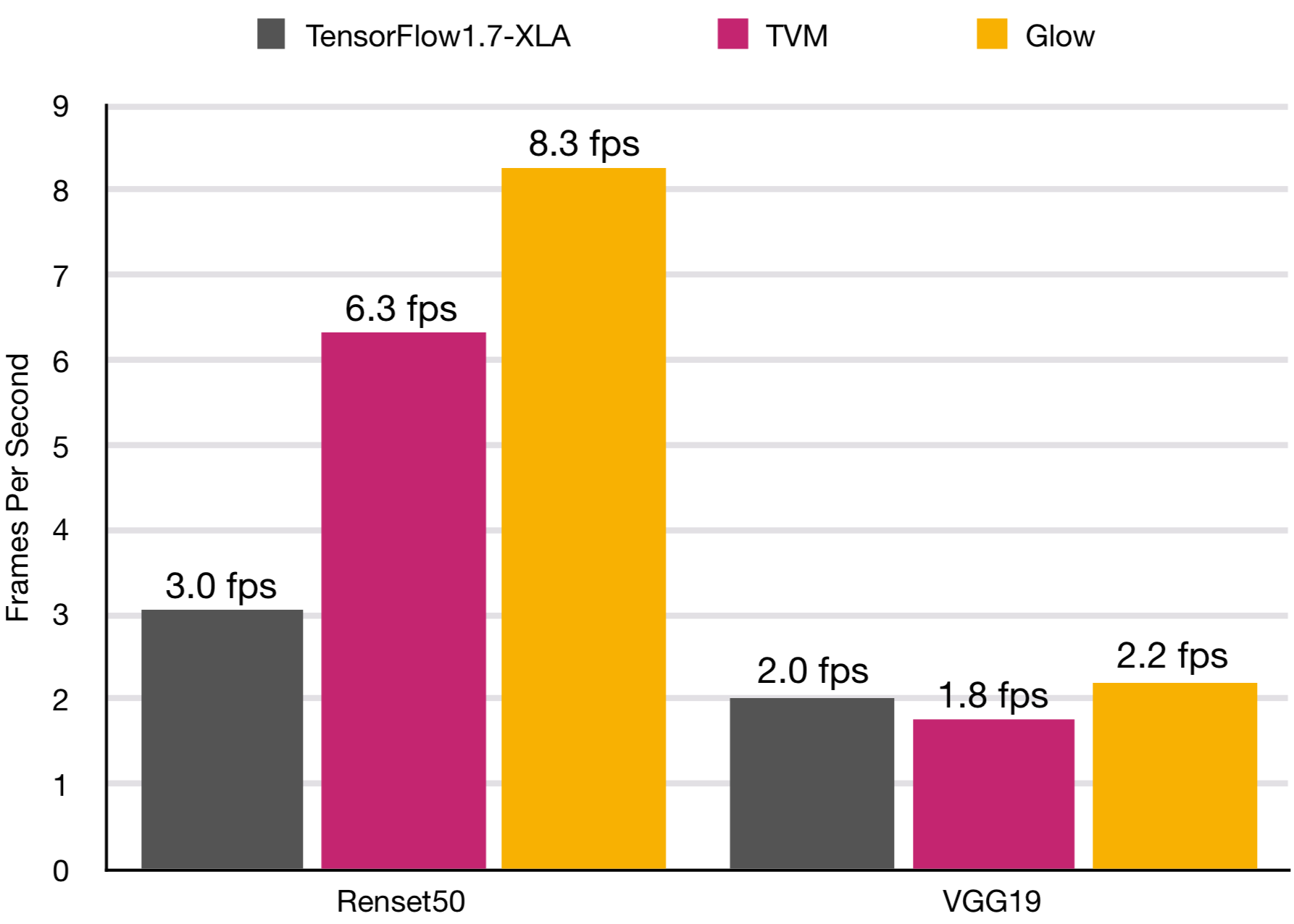}
\caption{Glow vs. TensorFlow-1.7 and TVM on an
  Intel\textsuperscript{\sffamily{\textregistered}} Core i7-7600U; frames per
  second on a single thread.}
\label{fig:glow_vs_tf}
\end{figure}

Two popular convolutional neural networks, Resnet50~\cite{resnet} and
VGG19~\cite{vgg}, are evaluated as seen in
Figure~\ref{fig:glow_vs_tf}. TensorFlow was compiled with XLA enabled. TVM was
compiled with LLVM 6.0, but without auto-tuning enabled or any specialized
schedules. Presented results used a batch size of 8 on all three
frameworks. Each framework's performance (in frames per second) did not vary
significantly across batch sizes of 2, 4, and 8.

As seen in Figure~\ref{fig:glow_vs_tf}, Glow is up to 2.7x faster than
TensorFlow. This is due to the fact that TensorFlow calls into Eigen which
implements convolution using the classic im2col followed by matrix
multiplication, while Glow compiles direct convolution
(Section~\ref{sec:resnet50_cpu}) and thus avoids im2col overhead. In addition,
Glow performs shape-aware code-generation.

Additionally, Glow is up to 1.3x faster than TVM. Note that we did not use
autotuning and improved schedules with TVM; we expect this would improve TVM's
performance. TVM does not use im2col like TensorFlow; similar to Glow, TVM
lowers nodes into a low-level IR. This IR is Halide-based, which generates LLVM
IR that is compiled for the native architecture. TVM and Glow also both generate
code with efficient memory access patterns such as tiling.

\section{Conclusion}
\label{sec:conclusion}

This paper presented the design of Glow, a machine learning compiler for
heterogeneous hardware. Glow lowers the compute graph of neural networks to
multi-level strongly-typed intermediate representations, enabling analyses and
optimizations appropriate for each level to efficiently and scalably target many
backends. We hope our efforts will enable research in the area of machine
learning acceleration.

\section{Acknowledgements}
In addition to the core Glow team, many fellow people at Facebook have made
contributions to the project, including Andrew Adams, Michel Aoun, William
Arnold, Sarah Bird, Brad Cottel, Stephen Chen, Evan Cheng, Soumith Chintala, Sy
Choudhury, Chris Dewan, Utku Diril, Marat Dukhan, Oniel Duncan, Dmytro
Dzhulgakov, Peter Goldsborough, Chris Gottbrath, Kim Hazelwood, Yangqing Jia,
Aravind Kalaiah, Daya S Khudia, Changkyu Kim, Bruce Lin, Howard Mansell, Erik
Meijer, Arun Moorthy, Sam Naghshineh, Maxim Naumov, Avinash Nayak, Pieter
Noordhuis, Joe Pamer, Joel Pobar, Yuri Putivsky, Lin Qiao, Vijay Rao, Martin
Schatz, Alexander Sidorov, Joe Spisak, Narayanan Sundaram, Andrew Tulloch, Mor
Tzur, Nicolas Vasilache, Adam Weis, and Hector Yuen.

We also would like to thank Eli Bendersky, Chris Leary, Richard Wei, and Tianqi
Chen for the development and release of their work to the open source.